\documentclass[vecphys]{svmult}

\usepackage{amssymb}
\usepackage{url}
\usepackage{makeidx}         
\usepackage{graphicx}        
\usepackage{multicol}        
\usepackage[bottom]{footmisc}

\makeindex             

\begin{document}

\title*{Coarse graining the Cyclic Lotka-Volterra Model: SSA and
    local maximum likelihood estimation.}
    \titlerunning{Coarse-graining SSA LLV models via local diffusion models}
\author{C. P. Calderon \inst{1}
 \and G. A. Tsekouras \inst{2,3} \and A. Provata \inst{2} \and I. G. Kevrekidis \inst{1,4} }
\institute{Department of Chemical Engineering, Princeton University,
Princeton, New Jersey 08544-5263, USA
\and Institute of Physical Chemistry,
National Research Center ``Demokritos'',
15310 Athens, Greece
\and Physics Department, University of Athens,
Panepistimioupolis, 10679 Athens, Greece
\and Corresponding Author: \texttt{yannis@arnold.Princeton.edu}}
%
%
\maketitle

\newcommand{\pe}{\psi}
\def\fo{{\mathcal{F}_{OP}}}
\def\d{\delta}
\def\ds{\displaystyle}
\def\e{{\epsilon}}
\def\ait{{A{\"\i}t-Sahalia }}
\def\aits{{A{\"\i}t-Sahalia's }}
\def\eb{\bar{\eta}}
\def\enorm#1{\|#1\|_2}
\def\Fp{F^\prime}
\def\fishpack{{FISHPACK}}
\def\fortran{{FORTRAN}}
\def\gmres{{GMRES}}
\def\gmresm{{\rm GMRES($m$)}}
\def\Kc{{\cal K}}
\def\norm#1{\|#1\|}
\def\wb{{\bar w}}
\def\zb{{\bar z}}


\def\bfE{\mbox{\boldmath$E$}}
\def\bfG{\mbox{\boldmath$G$}}

\begin{abstract}

When the output of an atomistic simulation (such as the Gillespie
stochastic simulation algorithm, SSA) can be approximated as a
diffusion process, we may be interested in the dynamic features of
the deterministic (drift) component of this diffusion.
We perform traditional scientific computing tasks (integration,
steady state and closed orbit computation, and stability analysis)
on such a drift component using a SSA simulation of the Cyclic Lotka-Volterra system
as our illustrative example.
The results of short bursts of appropriately initialized SSA
simulations are used to fit local diffusion models using \aits
transition density expansions \cite{ait2,aitECO,aitVEC} in a
maximum likelihood framework.
These estimates are then coupled with standard numerical algorithms
(such as Newton-Raphson or numerical integration routines) to help
design subsequent SSA experiments.
A brief discussion of the validity of the local diffusion approximation of the SSA simulation (a jump
process) is included.

\end{abstract}

\section{Introduction}
Reactive particle dynamic models arise in scientific fields
ranging from physical and chemical processes to systems biology
\cite{nicolis1,nicolis2,yanniscatalyst,VAguy,doyle,chemlang,NatureCircRhythm}.
Incorporating successive levels of detail in the modeling quickly
leads to models that are analytically intractable, necessitating
computational exploration.
Gillespie's  Stochastic Simulation Algorithm (SSA) and its variants \cite{VAguy,tau2,chemlang}
have gained popularity in recent years for modeling so-called mixed reacting systems; the approach
provides a middle ground between detailed molecular dynamics and lumped, Ordinary Differential Equation (ODE)
descriptions of chemical kinetics, incorporating fluctuations.
Knowing the kinetic scheme underlying such a simulation allows one to write,
at the infinite particle limit, the corresponding kinetic ODE.
%
At intermediate particle numbers ($N_{mol}$), the SSA has been approximated
with the continuous
``chemical Langevin equation" \cite{chemlang}.

In what follows we will assume that the results of an SSA simulation can
be successfully approximated through a continuous diffusion process.
Explicit knowledge of the drift and noise components of
such a process allows one to easily analyze certain features of the overall
behavior; one might, for example, be interested in the bifurcation behavior
of the ``underlying" drift component of the model, including
 the number and  stability of its steady states and their parametric
dependence.
In our work we assume that the only available simulation tool is a
``black box" SSA simulator, in which the mechanistic rules have
been correctly incorporated, but which we, as users, do not know:
we can only observe the SSA simulator {\it output}.
We want to perform a quantitative computational study of the
underlying drift component.
Since we cannot derive it in closed form (not
knowing the evolution rules), we want to perform this study using the least
possible simulation with the SSA code.
The approach we use follows the so-called ``equation-free" framework \cite{CommMathSci,yannisgear}:
in this framework traditional numerical algorithms become protocols for designing short bursts
of numerical experiments with the SSA code.
The quantities necessary for numerical computation with the
unavailable model (time derivatives, the action of Jacobians)
are estimated locally by processing the ``fine scale" SSA simulations.
In this work we extract such numerical information via parametric
local diffusion models using the transition density expansions
proposed by \ait \cite{aitECO,aitVEC}.
The numerical procedures we illustrate can also be used, in principle,
for different types of ``fine scale" models if their output happens
to be well approximated by diffusion processes.

The article is organized as follows:  In Section \ref{s_lvmodel}
we describe our illustrative model system. We then quickly outline
the basic ideas underlying equation-free numerics (Section
\ref{s_eqfreecomp}), and discuss our estimation procedure (Section
\ref{s_estproc}).
Our computational results are presented in Section
\ref{s_illofeqfreecomp}, and we conclude with a discussion
including goodness-of-fit issues.

\section{The Lattice Lotka-Volterra Model}
\label{s_lvmodel}


Our Cyclic Lotka-Volterra \cite{LLV,durrett}
illustrative example consists of a three-species ($X$, $Y$ and $S$)
nonlinear kinetic scheme of the following form \cite{LLV}:

\begin{eqnarray}
\label{LV_SSA}
\nonumber X \,+ \,Y \quad & &\mathop{\rightarrow}^{k_1} \quad 2Y \\
\nonumber Y \,+ \,S   \quad & &\mathop{\rightarrow}^{k_2} \quad 2S \\
\nonumber S \,+ \,X   \quad & &\mathop{\rightarrow}^{k_3} \quad 2X.
\end{eqnarray}

In the remainder of the paper we will refer to it simply as LV. In the deterministic limit, this kinetic scheme gives rise to a
set of three coupled nonlinear ODEs for the evolution of the
concentrations $X,Y$ and $S$.

\begin{eqnarray}
 {dX\over dt}&=&-k_1XY+k_3XS \\
\nonumber {dY\over dt}&=&-k_2YS+k_1YX \\
\nonumber {dS\over dt}  &=&-k_3SX+k_2SY \label{mfeq}
\end{eqnarray}

The total concentration ($X+Y+S$) is constant over time; setting
(without loss of generality) this constant to unity and
eliminating $S$
\begin{eqnarray}
\nonumber X+Y+S=1,\>\>\> \Longrightarrow \>\>\> S=1-X-Y
\label{eqno3}
\end{eqnarray}
reduces the system to

\begin{eqnarray}
\label{LV_ode}
{dX\over dt}&=&X\left[k_3-k_3X-(k_1+k_3)Y\right]\\
\nonumber {dY\over dt}&=&Y\left[-k_2+(k_1+k_2)X+k_2Y\right].
\end{eqnarray}

For every (positive) value of $k_1, k_2$ and $k_3$
four fixed points exist: three trivial and one non-trivial
steady state:

\begin{eqnarray}
\nonumber X_{s} &=& 0,\, Y_{s}=0,\, S=1 \quad(\rm {system\, invaded\, by\,} S) \\
\nonumber X_{s} &=& 1,\, Y_{s}=0,\, S=0 \quad(\rm {system\, invaded\, by\,} X) \\
\nonumber X_{s} &=& 0,\, Y_{s}=1,\, S=0 \quad(\rm {system\, invaded\, by\,} Y) \\
\nonumber X_{s} &=& {k_2\over {K}},\, Y_{s}={k_3\over {K}},
\nonumber S_{s} = {k_1\over {K}} \quad (\rm {nontrivial\, fixed\,
point})
\end{eqnarray}

where
\begin{eqnarray}
K=k_1+k_2+k_3.
\label{eq51}
\end{eqnarray}

An interesting feature of the phase space of the deterministic
model is the existence of a one-parameter family of closed orbits
surrounding a ``center'' (see Figure \ref{f_NRorbIllustration}).
The neutral stability of these orbits affects, as we will see below,
the fixed point algorithms used to converge on them.
The system is simulated through both the ODEs (\ref{LV_ode}) and
through an SSA implementation of the kinetic scheme (\ref{LV_SSA}) using $k_1,k_2=0.5$ and $k_3=0.7$ throughout.

\section{Equation Free Computation}
\label{s_eqfreecomp}
 The basic premise underlying equation-free
modeling and computation is that we have available a ``black box"
fine-scale dynamic simulator, and we believe that an {\it
effective} evolution equation exists (closes) for some set of
(coarse-grained) {\it outputs} or {\it observables} of the fine
scale simulation.
As discussed in more detail in  \cite{yannisgear,hummer},
one can numerically solve this (unavailable explicitly) equation through linking
traditional numerical methods with the fine scale code; in particular, the classical
continuum algorithms become protocols for the design of short, appropriately initialized
numerical experiments with the fine scale code.
The process starts by identifying the appropriate coarse-grained observables
(sometimes also called order parameters); typically these variables are low-order
moments of microscopically/stochastically evolving particle distributions
(e.g. concentrations for chemically reacting systems, like our example).
In general, good coarse-grained observables are not known, and data analysis
techniques to identify them from computational or experimental
observations are the subject of intense current research
\cite{Isomap,belkin,PNASCoifman}.
If the unavailable ``effective" equations are deterministic and
reasonably smooth, short runs of the fine-scale
simulator are used to estimate {\it time derivatives} of the
coarse-grained observables; initializing fine scale simulations
consistent with nearby values of the coarse-grained observables
gives estimates of directional derivatives (again assuming
appropriate smoothness), and can be linked with matrix-free
iterative linear algebra techniques (e.g. \cite{kelley}).
When an explicit evolution equation is available, these quantities, necessary in
numerical computation, are obtained through function or Jacobian evaluations of
the model formulas; here, they are estimated {\it on demand} from short computational
experiments with the fine scale solver.
If the underlying effective equation is stochastic, e.g. a diffusion,
then the results of the short simulation bursts must be used to estimate both the
drift and the noise components of the effective model - this is the case
we study here.
We will illustrate, using the SSA LV example, how
certain types of computations can be accelerated by appealing to classical numerical
methods.

\section{Estimation Procedure}
\label{s_estproc}
%
In what follows, we will assume that species concentrations are
good observables, and that the true process (the LV SSA simulation)
can be adequately approximated by a diffusion process, that is, a
stochastic differential equation (SDE) of the form:

\begin{equation}
 d\mathbf{X_t}=\mathbf{\mu(X_t;}\theta)dt+
 \mathbf{\Sigma(X_t;}\theta)
 d\mathbf{W_t}.
 \label{SDEgeneric}
\end{equation}
Here $\mathbf{X_t}$ is a stochastic process which is meant to
model the evolution of the observable(s),  $\mathbf{W_t}$
represents a vector of standard Brownian motions, and the
functions $ \mathbf{\mu(X_t;}\theta)$ and
$\mathbf{\Sigma(X_t;}\theta)$ are the drift and diffusion
coefficients of the process.
In the classical parametric setup, one assumes that the parameterized function
families to which the drift and diffusion coefficient functions belong
are known, and that the parameter vector $\theta$ is
finite dimensional.
In practice one rarely knows a
class of functions which can be used to describe the {\it global}
dynamics of the observables; in the equation free
computations below, however, we simulate the true process for only
relatively short bursts of time.
It therefore makes sense to (locally) consider the following SDE:

\begin{equation}
 d\mathbf{X_t}=\Big(\mathbf{A}+ \mathbf{B(X_t-X_o)}\Big)dt+\Big(\mathbf{C+
 D(X_t-X_o)}\Big)d\mathbf{W_t}.
 \label{eq_myaffine}
\end{equation}
where $\mathbf{W_t}$,$\mathbf{X_t}$,$\mathbf{X_o}$, $\mathbf{A}$
and  $\mathbf{C}$ $\in \mathbb{R}^d$ and $\mathbf{B}$ and
$\mathbf{D}$ $\in \mathbb{R}^{d \times d}$ (the $d$ Brownian
motions are assumed independent; the vector multiplying them, by
slight abuse of notation, contains the nonzero elements of the
diagonal matrix $\mathbf{\Sigma}$; extending to the correlated
case is straightforward).
This simple model is based on the fact that we expect smooth evolution of moments of
the observables, while at the same time taking into account the state dependence of the noise (neglecting
this dependence can cause bias in the estimation of the drift).
The parameters of this local linear model are estimated through
techniques associated with maximum likelihood estimation (MLE).
The motivation for using MLE techniques stems from the fact that
under certain regularity conditions \cite{vandervaart} such
estimators are (asymptotically) efficient as regards
the variance of the estimated parameter distribution.
In addition, the asymptotic parameter distributions associated
with MLE can sometimes be worked out analytically, or approximated
through Monte Carlo simulations; this knowledge can guide the
selection of the sample size necessary for a given desired
accuracy in coarse-grained computations \cite{llglassy}.

\subsection{Maximum Likelihood Estimation for Discretely Observed Diffusions}
We now recall a few basic facts about MLE estimation; standard
references include, e.g. \cite{hamilton,jeganathan,vandervaart}.
It is assumed throughout that the \emph{exact} distribution
associated with the parametric model admits a
continuous density whose logarithm is well defined
almost everywhere and is three times continuously differentiable with
respect to the parameters \cite{kl}.
%
%

MLE is based on maximizing the log-likelihood
($\mathcal{L}_{\theta}$) with respect to the parameter vector (for
our model $\theta \equiv
[\mathbf{A},\mathbf{B},\mathbf{C},\mathbf{D}]$):
\begin{equation}
\mathcal{L}_{\theta} \equiv \log\Big(f(\mathbf{x};\theta)\Big).
\label{eq_infoint}
\end{equation}

In the above equation, $\mathbf{x}$ corresponds to a matrix of
observations $\in \mathbb{R}^{d\times M}$ where $d$ is the
dimension of the state and $M$ is the length of the time series;
$f(\mathbf{x};\theta)$
corresponds to the probability of making observation
$\mathbf{x}$.
For a single sample path of a discretely observed diffusion known
to be initialized at $\mathbf{x}_0$, $f(\mathbf{x};\theta)$ can be
evaluated as \cite{hamilton}:

\begin{equation}
f(\mathbf{x};\theta)=\delta_{{x}_0}\prod \limits_{m=1}^{M-1}
f(\mathbf{x}_{m}|\mathbf{x}_{m-1};\theta).
\end{equation}

In this equation $f(\mathbf{x}_{m}|\mathbf{x}_{m-1};\theta)$
represents the conditional probability (transition density) of
observing $\mathbf{x}_{m}$ given the observation
$\mathbf{x}_{m-1}$ for a given $\theta$ and $\delta_{{x}_0}$ is
the Dirac distribution.
In our applications, we search for the parameter vector that is
best over {\it all} observations (we have an ensemble of $N$ paths
of length $M$).
In this case our expression for the log-likelihood (given the data
and transition density) takes the form:
\begin{equation}
\mathcal{L}_{\theta}:=\sum\limits_{i=1}^{N} \sum\limits_{m=1}^{M}
\log\Big(f(\mathbf{x}_{m}^i|\mathbf{x}_{m-1}^i;\theta)\Big).
\label{eq_loglikelihoodfuncdef}
\end{equation}
%

%
%

%
Assume the existence of an invertible symmetric positive definite
``scaling matrix" matrix $\mathcal{F}_{(M,\ \theta)}$
\cite{lecam00} associated with the estimator; the subscripts are
used to make the dependence of the scaling matrix on $M$ and
$\theta$ explicit. For the ``standard" case $N=1$ in time series
analysis, under some additional regularity assumptions
\cite{jeganathan,vandervaart}, one has the following limit for a
{\it correctly specified} parametric model:
\begin{equation}
\mathcal{F}_{(M,\
\hat{\theta})}^{\frac{1}{2}}(\theta_M-\hat{\theta})
\stackrel{\mathbb{P}_{\hat{\theta}}}{\Longrightarrow}
N(\mathbf{0},\mathbf{I}).
\end{equation}

Here $\hat{\theta}$ is the true parameter vector; $\theta_M$
represents the parameters estimated with a finite time series of
length $M$;
$\stackrel{\mathbb{P}_{\hat{\theta}}}{\Longrightarrow}$ denotes
convergence in distribution \cite{vandervaart,hamilton} under
$\mathbb{P}_{\hat{\theta}}$ (the distribution associated with the density
 $f(\mathbf{x};\hat{\theta)}$);
$N(\mathbf{0},\mathbf{I})$ denotes a normal distribution with mean
zero and an identity matrix for the covariance.
For a correctly specified model family,
$\mathcal{F}_{(M,\ \hat{\theta})}$ can be estimated in a variety of ways
\cite{white,lecam00}.
The appeal of MLE lies in that, asymptotically in $M$, the
variance of the estimated parameters
 is the smallest that can be achieved by an
estimator that satisfies the assumed regularity conditions
\cite{jeganathan,vandervaart}.

\subsection{Transition density expansions}
Here we briefly outline the key features of the recent work
of \ait \cite{aitECO,aitVEC} used in our coarse-grained computations below.
The problem with using even a simple model like that given in equation
\ref{eq_myaffine} is that the transition density associated with
the process is not known in closed form.
In recent years, many
attempts to approximate the transition density have
appeared in the literature; some techniques depend on analytical
approximations whereas others are simulation based (see, e.g.
\cite{ait2,aitECO,aitextension,bibby,gallant,smle}).
We have used, with some success, the expansions found in
\cite{ait2,aitVEC,aitECO}.
High accuracy can be obtained
using this method to approximate the transition density associated
with a {\it scalar} process; the multivariate case is discussed in \cite{aitVEC}.
The basic idea behind the scalar case, presented in \cite{ait2,aitECO}, is
as follows:
One first transforms the process given in equation
\ref{SDEgeneric} into a new process \cite{aitECO}:
%

\begin{eqnarray}
   & & dY_t= \mu_Y(Y_t;\theta)dt+ dW_t \\
  \nonumber & & Y\equiv \gamma(X;\theta)=\int^{X}\frac{du}{\sigma(u,\theta)} \\
 & & \lefteqn{\mu_Y(y;\theta)\equiv} \\ \nonumber  & &    \frac{\mu(\gamma^{-1}(y;\theta);\theta)}{\sigma(\gamma^{-1}(y;\theta);
\theta)}-\frac{1}{2}\frac{\partial \sigma}{\partial x}( \gamma^{-1}(y;\theta);\theta)
\end{eqnarray}

An additional change of variables brings the transition density of the process
closer to a standard normal density $Z \equiv
\Delta^{-\frac{1}{2}}(Y-y_o)$ where $\Delta$ is the time between observations.
The transformations introduced
allow the use of a Hermite basis set in order to approximate the
transition density of the original process via the following
series:

\begin{eqnarray}
 \lefteqn{p_Z(\Delta,z|y_o;\theta)\approx} \\
\nonumber & & \phi(z)
\sum\limits_{j=0}^{K}\eta_Z^{(j)}(\Delta,y_o;\theta)H_j(z)
\end{eqnarray}

\begin{eqnarray}
& & \lefteqn{\eta_Z^{(j)}(\Delta,y_o;\theta) \equiv} \\
\nonumber & & \lefteqn{\frac{1}{j!}\int\limits_{-\infty}^{\infty}H_j(z)p_Z(\Delta,z|y_o;\theta)dz:=} \\
\nonumber & &
\frac{1}{j!}\mathbb{E}[H_j\Big(\Delta^{-\frac{1}{2}}(Y_{t+\Delta}-y_o)\Big)|Y_t=y_o;\theta]
\end{eqnarray}

In the above, $H_j$ represents the $j^{th}$ Hermite polynomial and
$\phi(\cdot)$ is the standard normal density. The coefficients
needed for the approximation are obtained through the conditional
moments of the process $Y_t$.
\ait outlines \cite{aitECO} a procedure which exploits the
connection between the SDE and the associated Kolmogorov equations
in order to develop a closed form expression for the
$\eta_Z^{(j)}$ coefficients.
The approximation is exact if $K \rightarrow \infty$ and  the
coefficient functions satisfy the assumptions laid out in
\cite{aitECO}.
In numerical applications one must always deal with a finite K.
Problems may arise in the truncated expansion: the approximation
of the density may not normalize to unity or, worse, it may become
negative (see \cite{aitVEC,ait2,aitextension} for some possible
remedies).

In the multivariate case, it becomes more difficult to
introduce an analog of $Y_t$ \cite{aitVEC}.
Nonetheless, it is still possible to construct a series
motivated by the methodology used in the scalar case; however,
one now needs to expand in space and
time, whereas the Hermite expansion yielded a series ``in time
only" \cite{aitVEC}).
\ait \cite{aitVEC} outlines an approach which makes use
of a recursion for calculating the coefficients of the
expansion in the multivariate case.
We have had success in using these expansions, even in cases where
convergence of the infinite series is not guaranteed by the
conditions given in \cite{ait2,aitECO}.
Notice, for example, that our local models may allow a value of
zero for the diffusion coefficient; using a different function
class (made computationally feasible by the extension of Bakshi
and Ju \cite{aitextension,bakshi05}), such as sigmoidal functions
for it, may help circumvent such problems.
Other pathologies are discussed in \cite{llglassy}; estimates of
the range of the parameters of interest
\cite{lecam00,vandervaart,llglassy} can enhance the algorithm
performance. The comparison study \cite{jensen_poulsen} recommends
the use of the expansions by \ait for a wide class of diffusion
models.
Beyond the estimation itself, these expansions can also be helpful in
obtaining diagnostics that depend on knowledge of the transition density
(such as goodness-of-fit tests \cite{hong}) and asymptotic error analysis \cite{lecam00}.
%


\section{Illustrations of Equation-free Computation}
\label{s_illofeqfreecomp}

Having estimated the parameters of a local model at a given state point
opens the way to several computational possibilities.
Such estimates, for example, can be used in an iterative search for
zeroes of the (global, nonlinear) drift.
A Newton-Raphson iteration for a (hopefully better) guess of this
root involves the solution of set of linear equations for which
both the matrix and the residual are available from the local linear
drift.
The resulting estimate of the root is then used to launch a new
set of computational experiments with the ``inner" SSA code,
followed by a new estimation, linear equation solution, and so
on to convergence.
This illustrates the fundamental underpinnings of equation-free
computation.
Many numerical algorithms (here, root finding through Newton
iteration) do not really require good closed-form global models:
each iteration only requires local information (the first very few
terms of a Taylor series) in order to ``design" the next
iteration.
Traditional continuum numerical methods can thus be thought of
as {\it protocols} for the design of a sequence of model evaluations
(possibly model and Jacobian evaluations, occasionally even Hessian
evaluations).
In the absence of an explicit formula for the model, the same
protocol can be used to design {\it appropriately initialized
computational experiments} with a model of the system at a
different level (here, the SSA simulator).
Processing the results of these appropriately initialized short
bursts {\it estimate} the quantities required for scientific
computation, as opposed to {\it evaluating} them from a
closed-form model.
The so-called ``coarse projective integration" is another example
of the same principle.
Traditional {\it explicit} integration routines require a call to
a subroutine that {\it evaluates} the time-derivative of a dynamic model
at a particular state.
In the absence of an explicit model, short bursts of simulations
of a model of the system at a different level (again, here, SSA)
can be used to {\it estimate} these time derivatives, and, through
local linear models, extrapolate the state at a later time.
The fundamental assumption underpinning this entire computational
framework, is that an explicit evolution equation exists, and closes,
in terms of the (known) coarse-grained observables of the
fine-scale simulation (here, the concentrations of the SSA species).
If this, unavailable in closed-form, equation is {\it
deterministic}, then one only need to estimate a drift term from
fine scale simulations; if, on the other hand, the coarse-grained
equation is {\it stochastic} (fluctuations are important), then
both the local drift and diffusion terms must be estimated.
Certain computational tasks for stochastic {\it effective},
coarse-grained models require evaluations of {\it both} these
terms (e.g. computations of stationary, equilibrium densities, or
Kramers' type computations of escape times for bistable systems,
see for example \cite{dima,hummer}).
In this paper, we perform equation-free tasks for only
the drift component of the model; sometimes it may be interesting to
know whether the drift component dynamics possess zeros or closed loops,
as well as their parametric dependence.
Also, at infinite system size (practically, for sufficiently
large particle numbers) the SSA actually closes as a deterministic
ODE.
%


\subsubsection{Coarse Newton-Raphson for the fixed point of the drift.}
In what follows we work at system sizes large enough that a
diffusion approximation of the SSA output is meaningful, and
-even more- the dynamics of the drift component of the diffusion
are close to the kinetic ODE scheme dynamics.
The neutral stability of the fixed point and the closed loops
of the kinetic ODE suggest comparable features for the estimated
drift, which we set out to investigate.
We find the nontrivial root of the estimated drift
$\mathbf{F(X;}\theta)=\mathbf{0}$\footnote[1]{We use $\mathbf{F}(
\cdot;\theta)$ to denote the right hand side of a general
deterministic ODE; here $\mathbf{F}( \cdot;\theta)$ is the
estimated $\mathbf{\mu}(\cdot ;\theta)$} through a coarse
Newton-Raphson procedure as follows:
An ensemble of $N_{path}$ SSA simulations are initialized
in a neighborhood of the current guess $\mathbf{X}_0$ of the root.
Each is evolved in time, and the simulations
are sampled uniformly $M$ times during a time interval
of length $\tau$.
A local SDE model of the type (\ref{eq_myaffine}) is estimated using
the transition density expansions of \ait in an MLE-type
scheme; the resulting model parameters are used to
update the root guess through
\begin{equation}
\mathbf{X}_n=\mathbf{X}_{n-1}-\frac{\partial \mathbf{F}(\mathbf{X};\theta)
}{\partial
\mathbf{X}}^{-1}|_{\mathbf{X}=\mathbf{X}_{n-1}}\mathbf{F}(\mathbf{X}_{n-1};\theta)
\approx \mathbf{X}_{n-1}-\mathbf{B}^{-1}\mathbf{A}.
\end{equation}

\begin{figure} [h]
\centering
\includegraphics[angle=0,scale=.65]{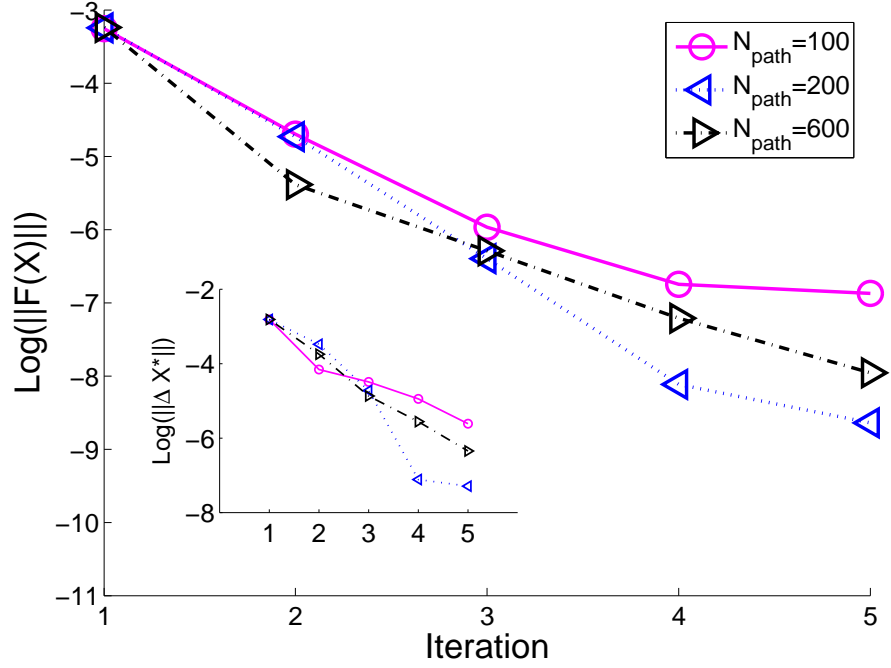}
\caption{Coarse NR to find stationary points.
Roots of $\mathbf{F(X)}\equiv \frac{\mathbf{dX}}{dt}$ using estimated (local) linear SDEs.
Parameters: $N_{mol}=1\times 10^4$, $\tau=5.032928126 \times
10^{-1},M=300 $. $N_{path}$ values are shown in the legend and the
$l^2$ distance of the current guess from the deterministic ODE
root is shown in the inset.  $\Delta \mathbf{X}^*$  (inset y-axis)
represents the difference between the current guess and the steady state of the ODE. }

\label{f_NRfp}
\end{figure}

Figure \ref{f_NRfp} shows this procedure for two different values of $N_{path}$
(other parameters are noted in the caption).
Newton-Raphson type procedures for isolated roots are known to converge
quickly given a good initial guess; furthermore, upon convergence,
the eigenvalues of the linearization of the drift are contained in
the matrix $\mathbf B$.
Estimates of these eigenvalues for different $N_{path}$ are
listed, upon convergence of the root finding procedure,  in Table
\ref{evalNR}.
The equation-free iterates approach the deterministic ODE root
(see inset); the latter is known to possess two pure imaginary
eigenvalues. 
The estimated (from local models) eigenvalues are also characterized by a relatively
small $(O(10^{-2}))$ real part.

\begin{table}
\caption{Representative real and imaginary parts of the eigenvalues of
the estimated drift upon convergence to the nontrivial fixed point
$\mathbf{X}\approx(0.2941,0.41176)$; the deterministic ODE
solution has a pair of pure imaginary eignenvalues. }

\begin{center} \footnotesize
\label{evalNR}
\begin{tabular}{|l|c|c|} \hline
& Re &Im \\ \hline
 $N_{path}=100$ &  $-2.53\times 10^{-2}$ &  $3.01 \times
10^{-1}$ \\ \hline $N_{path}=600$ &    $-2.39\times 10^{-2}$ &
$2.63 \times 10^{-1}$\\ \hline

\end{tabular}
\end{center}
\label{tab1}
\end{table}

\subsubsection{Coarse Projective Integration for the drift}
\begin{figure} [h]
\includegraphics[angle=0,scale=.65]{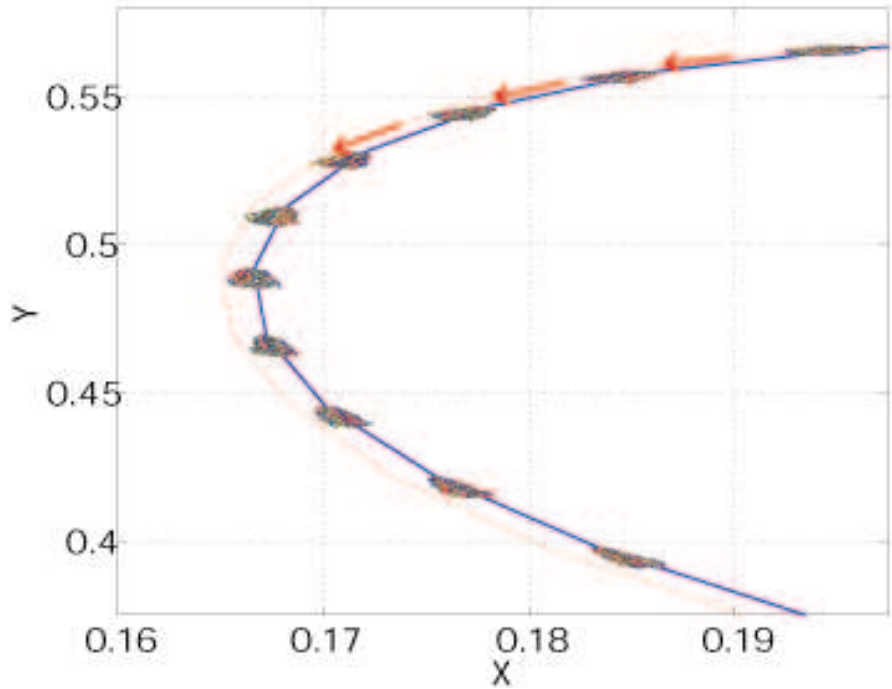}
\caption{Illustration of Coarse Projective Integration.
$N_{path}=600$, $N_{mol}=1\times10^4$, $\Delta t=.50329$,
$\tau=\frac{\Delta t}{4}$.} \label{f_varproint}
\end{figure}

A variety of numerical integration algorithms can be implemented in our framework.
Single step methods of the general form
\begin{equation}
\mathbf{X}_n=\mathbf{X}_{n-1}+
\mathbf{\Phi(X}_{n-1},\mathbf{X}_{n}; \Delta t).
\end{equation}
include the explicit and implicit Euler algorithms, (for which
$\mathbf\Phi$  is $\Delta t \mathbf{F(X}_{n-1})$ and $\Delta t
\mathbf{F(X}_{n})$ respectively).
Estimates of the drift at $\mathbf X_0$ can be immediately
used in a ``coarse forward Euler", while the estimated $\mathbf B$
can be used in a root-finding procedure, along the lines illustrated
above, in a ``coarse backward Euler" scheme.
Other schemes can be simply implemented.
Here we only demonstrate (explicit) coarse forward Euler;
predictor-corrector schemes (more appropriate for stiff problems)
are illustrated in  \cite{llalber}.
Representative results for our LV problem are shown in Figure
\ref{f_varproint}.
The deterministic ODE trajectory (dashed lines connecting points) is
compared to the projective integration of the drift component of
an SDE estimated locally from SSA simulation ensembles.
One clearly sees the evolution of the ensemble of SSA trajectories
initialized at every numerical integration point; $N_{path}$ such
trajectories were evolved and observed uniformly $M$ times over a
time interval $\tau$.
The results were processed through the estimation scheme and the
value of the drift at the original point $\mathbf X_0$ provided the
forward Euler estimate of the ``next" point through
$\mathbf{X}_1=\mathbf{X}_{0}+\Delta
t \mathbf{F}(\mathbf{X}_{0})$.
The procedure is then repeated.

Several algorithmic parameters must be carefully selected in such computations.
In our case the ``lifting" problem (the initialization of SSA simulations
at a given value of the coarse observables) is straightforward because of the
``mixed" nature of the SSA simulation; in general, the successful initialization
of a fine scale code consistent with a few coarse observables can be a complicated
and difficult issue, requiring, for example, preparatory constrained dynamic runs
\cite{Berendsen,Torrie}.

Another important parameter is the length of the integrator ``projective" step,
{\it $\Delta t$}, which for deterministic problems is set by stability and
accuracy considerations.
Stability discussions for projective integration can be found in \cite{CommMathSci};
here the issue is complicated by the fact that the model is {\it estimated}
rather than evaluated.
Multiscale methods for SDEs, including error estimates, can be
found in the work of \cite{Eijnden,ELiuEijnden}.
The total ``microscopic integration time" denoted by $\tau$ and
the time between observations $\equiv \delta t :=\frac{\tau}{M}$
also require careful selection.
If $\tau$ is too large, the simple linear model may break down as nonlinearities
in the real system manifest themselves.
If the assumed diffusion model is correct, there is no upper limit on $M$; yet a
diffusion approximation of a different underlying process, such as the
jump SSA here, will break down if the data is sampled too frequently.  Similar issues have been addressed in the control literature \cite{twoscale}.
Later on we will outline a goodness-of-fit test that can
be used to guide the selection of such algorithmic parameters.
In this work, short SSA trajectories in each ensemble
are initialized at the same base point $\mathbf X_0$,
or uniformly in a small neighborhood around it; we have not yet
explored optimal initialization.

\subsubsection{Equation-Free Coarse Variational Calculations}
A slight extension of the above coarse integration procedure
is the implementation of equation free integration of
{\it variational} equations.
The need for these arises naturally in our example when we attempt
to construct an algorithm that searches for possible closed orbits in the
dynamics of the estimated drift, and attempts to converge on them.
Closed orbits that are limit cycles can be found as (isolated)
fixed points of an appropriate Poincar{\'e} map.
In the deterministic LV problem, however, one has a one-parameter family
of such orbits, and the fixed points of the Poincar{\'e} map are not
isolated.
Anticipating a family of such closed orbits for our estimated drift model,
we {\it isolate} a single orbit from this one-parameter family by selecting
its period (the Poincar{\'e} return time).

\begin{figure} [h]
\includegraphics[angle=0,scale=.65]{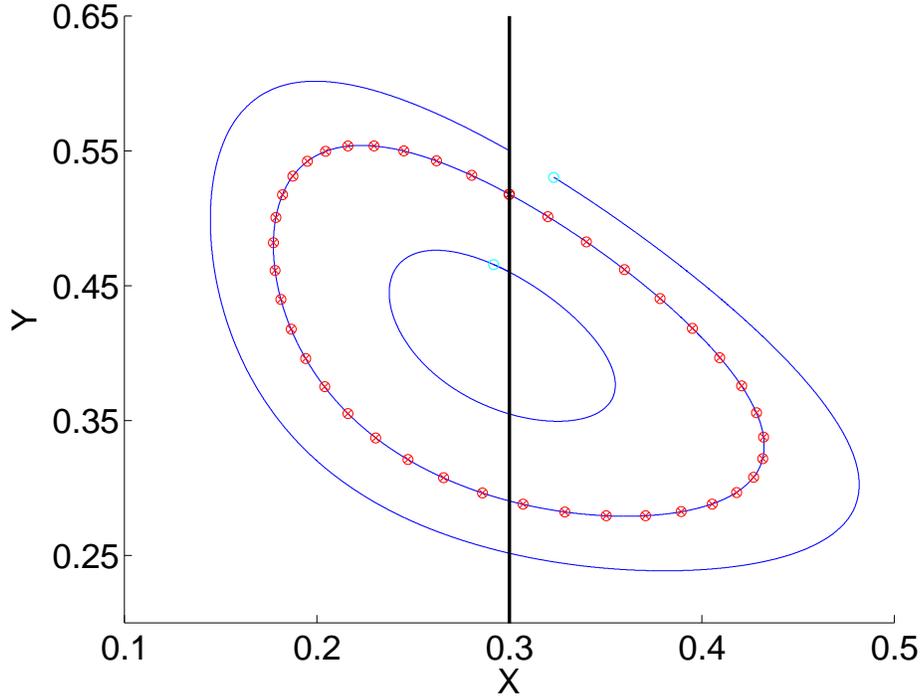}
\caption{Coarse closed orbit computations for the Lotka Volterra model.
The deterministic model phase portrait contains an infinite number of closed orbits.
Three such deterministic orbits (obtained by Runge-Kutta integration) are plotted here.
To find the closed orbit with a specified period $\tau$, we use the
Poincar\'{e} surface $X_P = 0.3$, shown as a solid line.
The Jacobian of the coarse Newton-Raphson scheme is computed through
variational integrations based on the estimated drift from ensembles
of SSA simulations initialized at the $N_{grid}$ base points shown
(see text). }

\label{f_NRorbIllustration}
\end{figure}

For a deterministic model, the initial value problem for the variational
equations is
\begin{eqnarray}
 & & \frac{d\mathbf{X}}{dt}=\mathbf{F(X;}\theta)  \\
 \nonumber & &  \mathbf{X}(t=0)=\mathbf{X^{IC}}  \\
   \nonumber & & \frac{d\mathbf{V}}{dt}=\frac{\partial \mathbf{F(X;}\theta)}{\partial \mathbf{X}}\cdot
\mathbf{V} \\ \nonumber & &
   \mathbf{V}(t=0)=\mathbf{I}.
 \label{ODEVARlim}
\end{eqnarray}
If $\mathbf{X}\in \mathbb{R}^d$ then $\mathbf{V} \in
\mathbb{R}^{d\times d}$. We use the results of integrating such
variational equations to locate closed orbits as zeroes of the
equation $\mathbf{G(X)\equiv X-\Phi_{\tau}(X)}$ where
$\mathbf\Phi_{\tau}(\cdot)$ represents the result of integration
from the (deterministic) initial condition $\mathbf X^{IC}$ for
time $\tau$.
To isolate the zeroes we seek, we select a Poincar{\'e} plane
through the value $X_P = 0.3$ of the first coordinate,
and the return time; we thus have one equation with one unknown,
the $Y$ coordinate of the intersection of our particular closed
orbit with the chosen Poincar{\'e} plane.
For our coarse integration, the return time $\tau$ is typically
too large to permit a single local diffusion model to accurately describe
the dynamics; we therefore use the following procedure:

\begin{itemize}
    \item Specify $\tau$ and the number $N_{grid}$ of local models we will use
    along the orbit, each valid for $T_f^{macro} :=\frac{\tau}{N_{grid}}$.
    \item Simulate $N_{path}$ SSA trajectories starting at the current fixed
        point guess; use the data as above to estimate the first local
        linear model. Use its drift (and the matrix ${\mathbf B}$) to obtain
        the next ``base point" as well as to step the variational equations
        for time $T_f^{macro}$.
    \item Repeat $N_{grid}$ times (see Figure \ref{f_NRorbIllustration}).
\end{itemize}

\begin{figure} [h]
\includegraphics[angle=0,scale=.65]{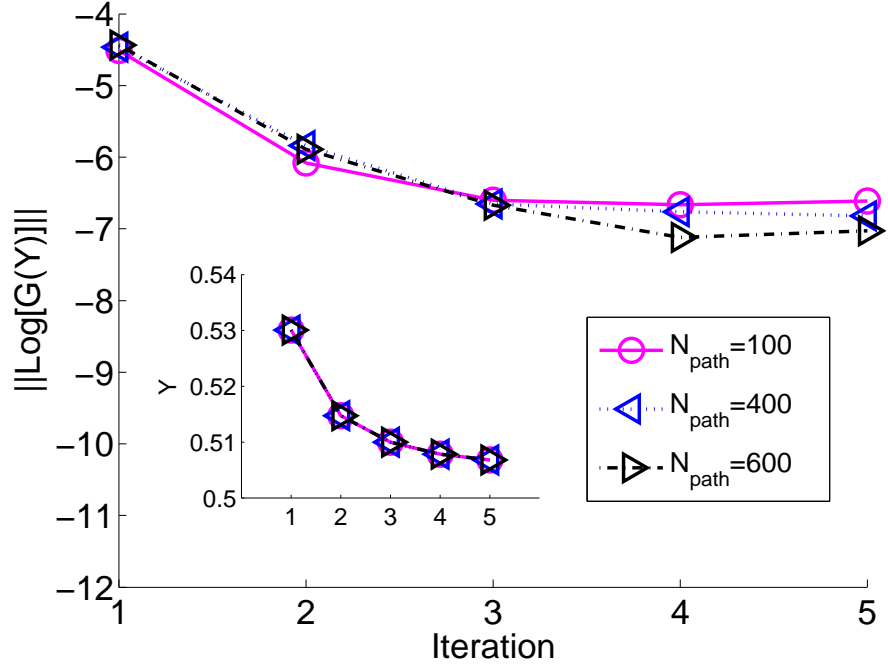}
\caption{Coarse Newton-Raphson for finding closed orbits of a specified period.
The zeroes of $G(Y)\equiv Y -\Phi_\tau(Y)$ were calculated
using a Jacobian evaluated from coarse variational integration based on
SSA simulations.
Parameters: $N_{mol}=1\times 10^4$, $\tau=2.0131712504 \times 10^{1}, M=300 $,
$N_{grid}=40$ ($N_{path}$ given in the legend).
The initial guess was $Y=0.53$.
For the deterministic ODE model the fixed point is $Y^{ODE}\approx 0.518$;
the coarse fixed point for $N_{path}=400$ was calculated to be $Y^{SSA}\approx 0.5075$.
 }

\label{f_NRorb_meth2}
\end{figure}

The output of this procedure gives us the residual of the fixed point equation
we wish to solve; the results of the variational integration at time $\tau$
(which, upon convergence, will give us an estimate of the monodromy matrix)
are then used to compute the Jacobian of the fixed point scheme.
One Newton-Raphson step for the $Y$ coordinate of the fixed point is taken,
and the procedure is then repeated.
Representative numerical results are shown in Figure
\ref{f_NRorb_meth2}.
Because of the neutral dynamics, the eigenvalues of the monodromy
matrix upon convergence are {\it both} equal to 1 (in the
deterministic ODE).
Table \ref{tab2} shows representative eigenvalue upon convergence
for different $N_{path}$ (sometimes the eigenvalues are numerically found
as complex conjugates with a small imaginary part).
Clearly, in addition to the algorithmic parameters involved in coarse projective
integration, we should now also take into account the desired accuracy of the
variational integration (quantified in part by the existence of an eigenvalue
equal to unity upon convergence).

\begin{table} [h]
\caption{Representative monodromy matrix eigenvalues upon convergence of the fixed point
iteration for two distinct $N_{path}$ computations (see text).}
\begin{center} \footnotesize
\begin{tabular}{|l|c|c|} \hline
$N_{path}=100$ & $(0.9698,0.1707)$ & $(0.9698,-.1707)$  \\ \hline
 $N_{path}=400$ & $(0.9159,0.0252)$ & $(0.9159,-.0252)$ \\ \hline

\end{tabular}
\end{center}
\label{tab2}
\end{table}

\section{Discussion}
\label{s_discussion}
We have illustrated the implementation of certain coarse-grained computations with the
LV model; one focus was the coarse-grained integration of the variational equations
for the SSA-based drift estimation, as well as the modifications of the coarse Newton-Raphson
iteration dictated by the neutral stability of the dynamics (the existence of infinitely
many closed orbits in the ODE limit, which appears to approximately persist in our
computations).
The second focus was the use of \aits expansions to estimate local linear
SDEs from short bursts of SSA data as an intermediate step.
This naturally leads to some crucial questions about the goodness-of-fit
of the simple SDE models: (a) is the diffusion approximation a ``good"
description of the dynamics?  (b) Is the linear approximation valid for
the time series length chosen? and (c) How reliable is the model for making
predictions/forecasts ?

\begin{figure}

\includegraphics[angle=0,scale=.65]{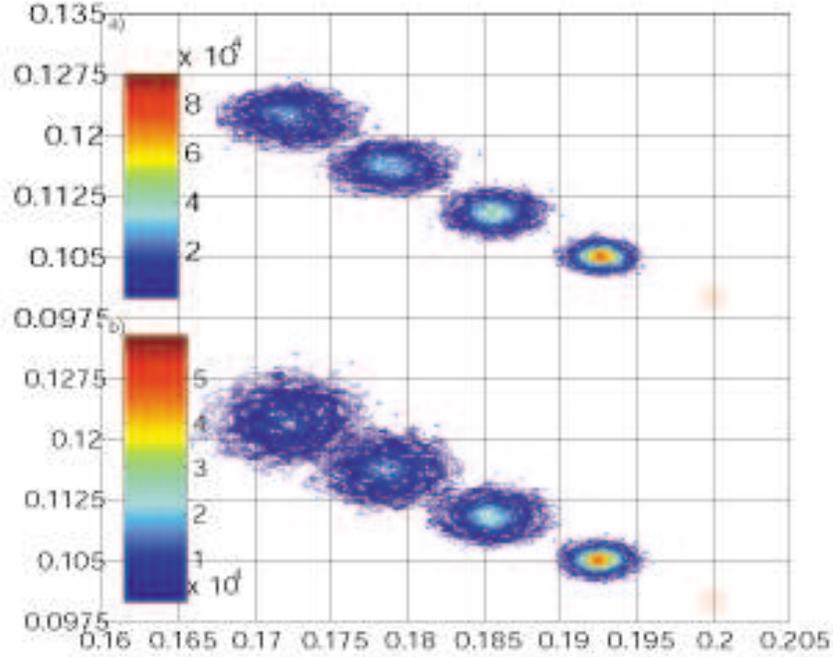}
\caption{Evolution of actual and model densities.
The top figure shows the evolution of the SSA process, initialized as a
Dirac distribution (marked in red);
the bottom plot shows an ensemble of numerical simulations of the ideal
diffusion model using the parameters estimated from the SSA.
Both distributions are plotted
at $\frac{\tau}{4},\frac{\tau}{2},\frac{3\tau}{4}$,and $ \tau$.
Relevant parameters: $N_{path}=5\times 10^3, M=300$, $N_{mol}=1\times10^4$,
$\tau=0.50329$.}
\label{edistevolve}
\end{figure}

\begin{figure}

\includegraphics[angle=0,scale=.65]{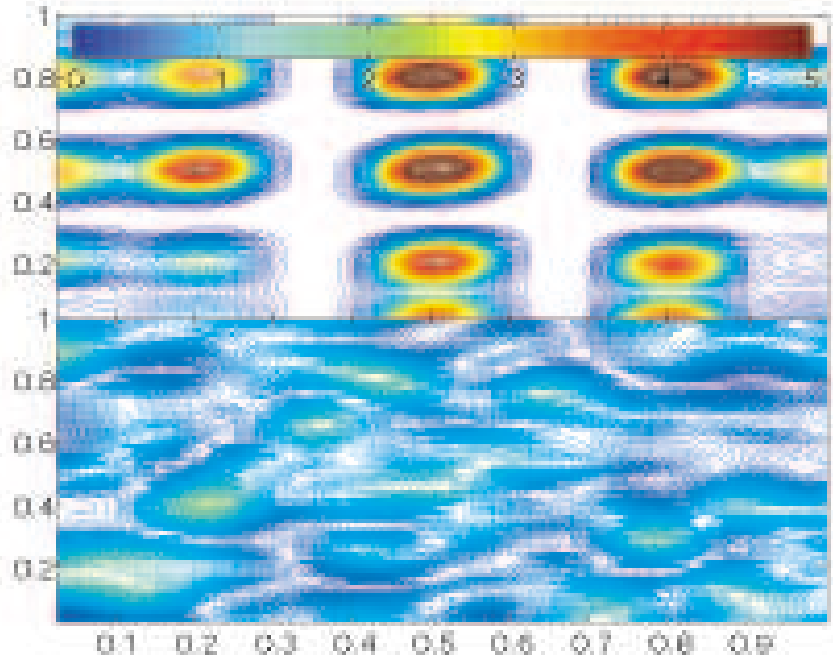}
\caption{Towards hypothesis testing. The function plotted corresponds
to an empirical estimate of the two-dimensional density
function described in \cite{diebold,hong}.
The data are obtained from the same ensemble of SSA simulations
as in Fig \ref{edistevolve}; the top figure is for $N_{mol}=1\times 10^4$ and
the bottom for $N_{mol}=4\times 10^5$.
In the infinite sample limit and for a correctly specified
model the density would be unity in the entire support ($[0,1]\times [0,1]$)  of the function.
The figure suggests that the observations of the larger system are
closer to a diffusion model. } \label{f_gof}

\end{figure}

One should quantitatively know how large $N_{mol}$ needs to be, for a
given sampling frequency, for a diffusion model to be a statistically
meaningful approximation \cite{kurtz05,chemlang}. Sampling too often may be detrimental in many diffusion approximations (e.g. \cite{twoscale}).
Local linear models (i.e. short truncations of Taylor series) are used
extensively in scientific computations, but only for short time steps,
whose length is determined by overall stability and accuracy considerations.
Similar considerations arise in choosing the $\tau$ used for
SSA data collection towards the estimation of the local linear SDE models used here;
clearly, when the underlying drift is nonlinear, $\tau$ cannot be too large.
%
%
A useful diagnostic tool for questions (a) and (b) applicable
if one does have an accurate transition density
approximation, is the probability integral transform
\cite{diebold,hong}.
Using the data and the (assumed known) exact transition density,
one creates a new random variable which,
for a correctly specified model, has a known distribution.
The method is applicable to both
stationary and non-stationary time series; furthermore it
depends on integrations of the transition density approximation rather
than differentiations.
Given the data, one (appropriately) estimates model parameters and
then constructs the random variables $Z_n$ for each observation
\footnote[2]{The method applies to both a vector and scalar process,
however the construction is easiest to demonstrate in the latter
case.  See \cite{hong} for the multivariate extension. } ($x_n$).  The construction below follows
that in Section 3 of \cite{diebold}:
\begin{eqnarray}
\nonumber Z_n:=\int\limits_{-\infty}^{x_n} p(x_n'|x_{n-1};\theta)dx_n' \\
\nonumber Z_n \sim q(Z_n)\equiv \frac{d \mathbb{Q}(Z_n)}{dZ_n}
\\ \nonumber
 x_n \sim f(x_n|x_{n-1})\equiv \frac{d \mathbb{F}(x_n|x_{n-1})}{dx_n}
\\ \nonumber
\end{eqnarray}
The symbol $\sim$ denotes that the random variable on the left of
the symbol is distributed according to the density to the right.
Under a correctly specified model, the $Z_n$'s  are independent
and uniformly distributed on $[0,1]$, independent of the
transition density \cite{diebold}.
In \cite{hong} a comprehensive suite of statistical tests are
reviewed which exploit knowledge of the transition density and the transformation shown above.
Figure \ref{f_gof} plots a kernel density estimate
 (see equation 6 on page 44 in \cite{hong}) which is based on the estimated
parameters and the observed data.
If the model is correctly
specified, the infinite sample size density should be the product of two uniform densities.
Test statistics can be created from this function (see \cite{hong}
for details).
%

Inspection of the figures shows that, for a particular
representative SSA ensemble run for $N_{mol}=1\times 10^4$, and a
particular sampling frequency, the diffusion approximation is not
acceptable; the situation appears better for $N_{mol}>4\times
10^5$.
It is interesting to notice that, while $N_{mol}=1\times 10^4$ is not
large enough for the conditions of Figure \ref{edistevolve}, visual inspection of the
empirical and the SDE-based density evolution might suggest otherwise.
In traditional, continuum numerical algorithms issues of on-line error estimation,
time-step and mesh adaptation are often built-in in modern, validated software.
There is a clear necessity for incorporating, in the same spirit, hypothesis
testing techniques in codes implementing the type of computations we
described here; yet automating such processes appears to be a major challenge.

\begin{figure}

\includegraphics[angle=0,scale=.65]{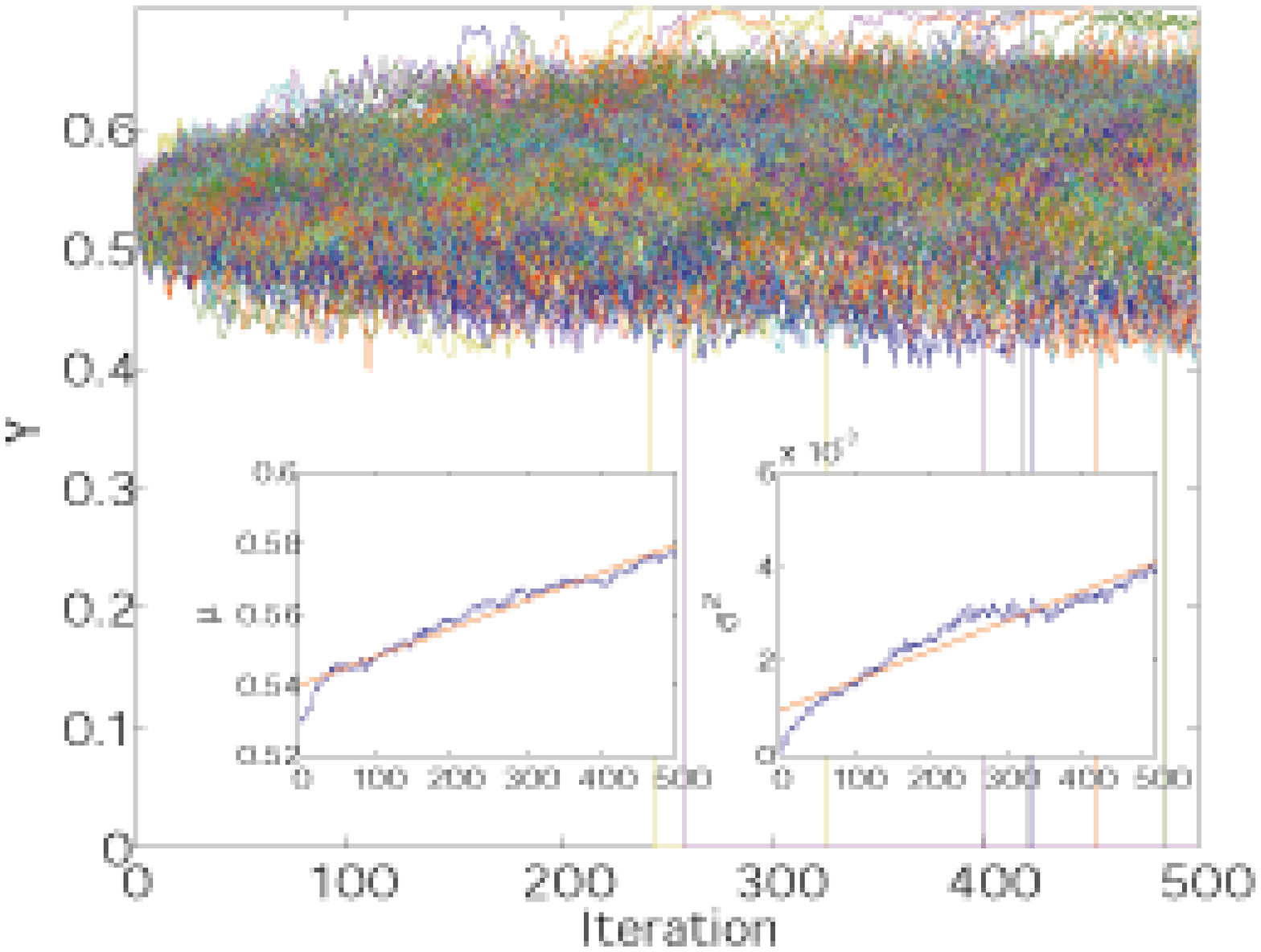}
\caption{SSA return map computations for the LV model.
An ensemble of 200 trajectories initialized at $\mathbf{X}=(0.3,0.53)$ are evolved,
and the $Y$ coordinate of their $X_P=0.3$ Poincar\'{e} map crossings in the
negative $X$ direction is recorded. The insets (see text) suggest a
systematic upward drift, bringing the system closer to extinction (this event is indicated by vertical lines, see text).
 }
\label{wandermapPATHS}
\end{figure}

In our next application, we evolved an ensemble of trajectories starting from a Dirac initial distribution, and then
recorded the Poincar\'{e} map for each individual trajectory over a long simulation
period.
Figure \ref{wandermapPATHS} shows the evolution of the $Y$ coordinate of these
trajectories as function of the map iterate.
For long times, different initial conditions in the ensemble approach some of the
``extinction" fixed points of the ODE vector field (see the vertical lines
in Fig. \ref{wandermapPATHS}); once there, the system no longer changes over time.
Visual inspection of the evolution of the ensemble suggests that
one might try to coarse-grain the Poincar{\'e} map evolution as a
model SDE; the insets in the figure show the initial evolution of
the mean and the variance of the Poincar{\'e} map iterates.
The smooth line in the insets, a simple least squares fit, seems to suggest a systematic
evolution towards ``larger" oscillations, bringing the system closer to extinction.
If this evolution could be well approximated locally by a
diffusion processes, approximations similar in spirit to the ones
shown in this article might be used to explore features of the
distribution of extinction times for the problem.

{\bf Acknowledgements} This work was partially supported by a Ford Foundation/NRC
Fellowship to (CC) and an NSF ITR grant (IK).


 \end{document}